\documentstyle[12pt,epsf]{article}  
\begin{document}    
\def\o{\over}
\def\Ar{\rightarrow}
\def\bar{\overline}
\def\r{\gamma}
\def\d{\delta}
\def\a{\alpha}
\def\b{\beta}
\def\n{\nu}
\def\m{\mu}
\def\k{\kappa}
\def\g{\gamma}
\def\r{\rho}
\def\s{\sigma}
\def\t{\tau}
\def\e{\epsilon}
\def\p{\pi}
\def\th{\theta}
\def\om{\omega}
\def\vp{{\varphi}}
\def\Re{{\rm Re}}
\def\Im{{\rm Im}}
\def\ra{\rightarrow}
\def\ti{\tilde}
\def\bar{\overline}
\def\l{\lambda}
\def\G{{\rm GeV}}
\def\M{{\rm MeV}}
\def\eV{{\rm eV}}
\setcounter{page}{1}
\thispagestyle{empty}
\topskip 2 cm
%\begin{flushright}    
%{NIIG-DP-00-0x \\,  October   2000}\\
%{hep-ph/}\\
%\end{flushright}   
\vskip 1 cm
\centerline{\Large \bf Large Mixing Angle MSW Solution}
\vskip 0.5 cm
\centerline{\Large\bf  in  $U(1)$ Flavor Symmetry}
\vskip 2 cm
\centerline{{\large \bf Morimitsu TANIMOTO}
  \footnote{E-mail address: tanimoto@muse.hep.sc.niigata-u.ac.jp} }
\vskip 1 cm
 \centerline{ \it{Department of Physics, Niigata University, 
 Niigata 950-2181, JAPAN}}
\vskip 3 cm
\centerline{\Large \bf Abstract}
\vskip 1 cm
 We have discussed the quark-lepton mass matrices with the $U(1)$ flavor
symmetry in SU(5), which lead to the large mixing angle MSW solution of solar 
neutrinos.
 The solar neutrino solution depends on  the
next-leading terms in the neutrino mass matrix.
We have found the lepton mass matrices with the $U(1)\times Z_2$
symmetry, which give the LMA-MSW solution uniquely.
 The coefficients of the matrix elements of  the charged leptons 
 are  constrained strongly due  to the bound of  $U_{e3}$
 in the  CHOOZ experiment.
%%%%%%%%%%%%%%%%%%%%%%%%%%%%%%%%%%%%%%%%%%%%%%%%%%%%%
\newpage
\baselineskip=24 pt
\topskip 0 cm
%%%%%%%%%%%%%%%%%%%%%%%%%%%%%%%%%%%%%%%%%%%%%%%%%%%%%%%%%%%%%%%%%%%%%%%%%%%%%%%
%%%%%%%%%%%%%%%%%%%%%%%%%%%%%%%%%%%%%%%%%%%%%%%%%%%%%%%%%%%%%%%%%%%%%%
 Super-Kamiokande  has almost confirmed  the neutrino oscillation
 in atmospheric neutrinos, which favors the $\n_\mu\Ar \nu_\tau$
process with  $ \sin^2 2\th_{\rm atm} \geq 0.88$ and 
 $ \Delta m^2_{\rm atm}=  (1.5\sim 5)\times  10^{-3} \eV^2$ \cite{SKam}.
For the solar neutrinos \cite{SKamsolar}, 
 the 1117 days data in Super-Kamiokande favors 
 the LMA-MSW solution \cite{N2000}, which is
  $\sin^2 2\th_\odot=0.65\sim 0.97$ and  
 $\Delta m_{\odot}^2= 10^{-5}\sim 10^{-4}\eV^2$.  However,
 there are still allowed four solutions,
   the small mixing angle (SMA) MSW \cite{MSW}, 
the large mixing angle (LMA) MSW, the low $\Delta m^2$ (LOW)  and  
the vacuum oscillation (VO) solutions \cite{BKS}.

 There are a lot of ideas for the single large mixing angle in the
 neutrino mixing matrix (MNS matrix) \cite{MNS}.
However,  it is not easy to get the nearly bi-maximal mixings \cite{bimax}
with the LMA-MSW mass scale in the GUT models \cite{GUT}.
Therefore, it is important  to search for  the texture of the lepton
mass matrix  with the LMA-MSW solution.

In this paper,
we study how to get the LMA-MSW solution with the help of 
the $U(1)$ flavor symmetry.
 Vissani \cite{Vissani} has already shown that the texture of the
neutrino mass matrix with the $U(1)$ flavor symmetry,
%%%%%%%%%%%%%%%%%%%%%%%%%%%%%
\begin{eqnarray}
 M_\n\sim   \left( \matrix{\e^2 & \e & \e \cr
           \e & 1 & 1 \cr  \e & 1 & 1 \cr  } \right) \ ,
\label{typical}
\end{eqnarray}
%%%%%%%%%%%%%%%%%%%%%%%%%%%%
 \noindent  leads to  the LMA-MSW solution
in the case of   $\e\simeq 0.05$.   
Recently, Sato and Yanagida  \cite{SY} have also studied this texture
numerically and showed clearly that both the  LMA-MSW and SMA-MSW solutions
are obtained. However, since they have not discussed the quark mass matrices,
 simple relations among the quark-lepton mass matrices are not obtained.
We discuss  the quark-lepton mass
matrix textures with the $U(1)$ flavor symmetry in the SU(5) GUT
focusing on the solar neutrino solutions.
After careful study in the  $U(1)$ symmetry,
we also propose  other textures of the quarks and leptons
 in the $U(1)\times Z_2$ symmetry.

%%%%%%%%%%%%%%%%%%%%%%%%%%%%%%%%%%%%%%%%%
%%%%%%%%%%%%%%%%%%%%%%%%%%%%%%%%%%%%%%%%%

Assuming that oscillations need only account for 
 the solar and the atmospheric neutrino data, we consider the 
LMA-MSW solution, in which 
the  mixing matrix  and the neutrino masses are given as

\begin{eqnarray}
U_{MNS}\sim  \left ( \matrix{\frac{1}{\sqrt{2}}& -\frac{1}{\sqrt{2}} & 0\cr
                \frac{1}{2} &  \frac{1}{2}& -\frac{1}{\sqrt{2}} \cr
 \frac{1}{2} & \frac{1}{2} &  \frac{1}{\sqrt{2}} \cr } \right ) \ ,
 \quad \qquad 
 \frac{\Delta m_{\odot}^2}{\Delta m^2_{\rm atm}}=0.01 \sim 0.1 \ .
\end{eqnarray}

\noindent 
 If  neutrino masses are  hierarchical, we expect the neutrino mass ratio
$m_{\nu 2}/m_{\nu 3}=\l^2 \sim \l$  with $\l\simeq 0.2$, which is
similar to the charged lepton mass hierarchy.
%%%%%%%%%%%%%%%%%%%%%%%%%%%%%%%%%%%%%%%%%%%%%%%%%%%%%
%%%%%%%%%%%%%%%%%%%%%%%%%%%%%%%%%%%%%%%%%%%%%%%%%%%%%%%%
%%%%%%%%%%%%%%%%%%%%%%%%%%%%%%%%%%%%%%%%%%%%%%%%%%%%%%%%

Let us consider the $U(1)$ flavor symmetry \cite{U1,Ramond}, in which
 fermions carry $U(1)$ charges, 
  $U(1)$  is spontaneously broken by  the VEV of 
  the elctroweak  singlet with $U(1)$ charge $ -1$, and
Yukawa couplings appear as effective operators
through Froggatt-Nielsen mechanism  \cite{FN}.
  When we integrate out massive fermions, the effective Yukawa couplings
 of the leptons below the mass scale $\Lambda$ are of the form:
 
  \begin{equation}         
    L_i \bar \ell_j H_d \left (  {S\o \Lambda} \right )^{m_{ij}} +
    \frac{1}{M_R} L_i  L_j H_u H_u \left (  {S\o \Lambda} \right )^{n_{ij}} \ ,
 \label{effectiveL}
  \end{equation}

   \noindent
  where $S$ is the singlet scalar of the SM, which breaks the flavor symmetry 
spontaneously by the VEV $<S>$, 
  $M_R$ is a  relevant high mass scale,
 and  $m_{ij}=a_i+b_j+h_d$ and $n_{ij}=a_i+a_j+h_u \ (i,j=1,2,3)$, in which
 $a_i$, $b_i$, $h_d$ and $h_u$ are  $U(1)$ charges for the  doublets
$L_i$,  the singlets $\bar \ell_i$, the Higgs $H_d$ and the Higgs $H_u$,
respectively.
%%%%%%%%%%%%%%%%%%%%%%%%%%%%%%%%%%%%%%%%%%%%%%%%
\footnote{In the seesaw mechanism \cite{seesaw}, the $U(1)$ charge of the
right-handed neutrinos should be assigned. However, in the effective Yukawa
couplings after integrating out the heavy righ-handed neutrinos,
their  $U(1)$ charges cancel as far as one uses the same
 ${S/\Lambda}$ for the right-handed Majorana couplings \cite{Ramond}. 
Therefore, we do not address their $U(1)$ charges in this paper.}
%%%%%%%%%%%%%%%%%%%%%%%%%%%%%%%%%%%%%%%%%%%%%%%
  We work in the supersymmetric model.
 In non-supersymmetric models, powers of $S^\dagger$ should be allowed.
  However, since  this possibility is forbidden in the super-potential
 of the supersymmetric model, we ignore $S^\dagger$.

%%%%%%%%%%%%%%%%%%%%%%%%%%%%%%%%%%%%%%%%%%%%%%%%%%%%
%%%%%%%%%%%%%%%%%%%%%%%%%%%%%%%%%%%%%%%%%%%%%%%%%%%%

 We discuss the LMA-MSW solution in the  SU(5) GUT.
Taking the U(1) charges of ${\bf 5^*}$ and ${\bf 10}$ fermions
 for the (1st, 2nd, 3rd) family \cite{AF}
 in order to get the LMA-MSW solution:
%%%%%%%%%%%%%%%%%%%%%%%%%%%%%%%%%%%%%%%%%%%%%%%%%%%%
\footnote{Our assignment of the $U(1)$ charges is  different from the
one in ref.\cite{AF}, where the LMA-MSW solution is  not obtained.
Our  assignment is unique to get the textute in eq.(\ref{typical})
 being consistent with  the quark sector. }
%%%%%%%%%%%%%%%%%%%%%%%%%%%%%%%%%%%%%%%%%%%%%%%%%%%%

\begin{equation}
\Psi_{\bf 5^*}\sim (A+2,\ A,\ A) \ ,\quad\qquad 
\Psi_{\bf 10} \sim (A+3,\ A+2,\ A)\ , 
\label{SU5}
\end{equation}
\noindent
 and putting 
\begin{equation}
  \frac{<S>}{\Lambda} = \l  \ , 
\end{equation}
\noindent we obtain  the quark mass matrices as follows:
%%%%%%%%%%%%%%%%%%%%%%%%%%%%%%%%%%%%%%%%%%%
\footnote{We take  $h_u=h_d=0$ and $A=0$ in our calculation.
These choices do not spoil our conclusion.}
%%%%%%%%%%%%%%%%%%%%%%%%%%%%%%%%%%%%%%%%%%%

\begin{eqnarray}
 M_U\sim   \left( \matrix{\l^6 & \l^5 & \l^3 \cr
                           \l^5 & \l^4 & \l^2 \cr
                           \l^3 & \l^4 & 1 \cr  } \right) \ , \qquad\quad
 M_D\sim
             \left (\matrix{\l^5 & \l^3 & \l^3 \cr
			    \l^4 & \l^2 & \l^2    \cr
                            \l^2 & 1 & 1    \cr } \right) \ . 
\end{eqnarray}

\noindent
These mass matrices are consistent with the experimental values
of the quark masses and the CKM mixing matrix \cite{CKM}, except for
the value of the lightest u-quark mass.

%%%%%%%%%%%%%%%%%%%%%%%%%%%%%%%
The lepton mass matrices are given as follows:

\begin{eqnarray}
 M_E\sim
             \left (\matrix{\l^5 & \l^4 & \l^2 \cr
			    \l^3 & \l^2 & 1    \cr
                            \l^3 & \l^2 & 1    \cr } \right) \ , \qquad\quad
 M_\n\sim   \left( \matrix{\l^4 & \l^2 & \l^2 \cr
                           \l^2 & 1 & 1 \cr
                           \l^2 & 1 & 1 \cr  } \right) \ .
\label{texture1}
\end{eqnarray}

\noindent
The left-handed mixings of the charged lepton and the neutrino
between the second and the third family are almost
maximal  $s_{23}^E\simeq s_{23}^\nu  \simeq 1/\sqrt{2}$.
Neglecting other mixings, 
  the MNS mixing matrix is written as:

\begin{eqnarray}
 U \equiv  L^\dagger_E L_\nu \simeq
 \left (\matrix{1 & 0 & 0 \cr
		  0 & 1/\sqrt{2} & 1/\sqrt{2} \cr
                  0 & -1/\sqrt{2} & 1/\sqrt{2}\cr  } \right)^T
 \left (\matrix{1 & 0 & 0 \cr
			    0 & e^{i\a} & 0 \cr
                            0 & 0 & 1 \cr  } \right) 
 \left (\matrix{1 & 0 & 0 \cr
			    0 & 1/\sqrt{2} & 1/\sqrt{2} \cr
                            0 & -1/\sqrt{2} & 1/\sqrt{2} \cr  } \right) \ ,
\label{max}
\end{eqnarray}

\noindent
where the phase matrix is needed as well as in  the quark sector
\cite{phase}
because  the lepton mass matrices are complex in general.
This unknown  phase is related with the CP violation in the lepton sector.
If $\a=\pm \pi/2$ is taken,  we  get $|U_{\mu 3}|=1/\sqrt{2}$, 
 which is still the maximal mixing. Thus, maximal mixings of both 
sectors lead to the nearly maximal mixing of the MNS matrix 
without cancellation if  the relevant phase is chosen.

%%%%%%%%%%%%%%%%%%%%%%%%%%%%%%%%%%%%%%%%%%%%%%%%%%%%%%

 On the other hand, the  left handed mixings  
between the first and the second (third) family should be carefully
examined. The left-handed mixings  in the charged lepton sector are  given as 
 \cite{HR}:
\begin{equation}
  s_{12}^E \simeq \frac{Y_{12}}{Y_{22}-Y_{23}Y_{32}}\ ,
\qquad 
  s_{13}^E \simeq Y_{13} \ ,
\end{equation}
\noindent where
$Y_{ij}$'s  are (i, j) components of the mass matrix with 
the normalization of $Y_{33}=1$.
Supposing that the accidental cancellation in the denominator of
$s_E$ does not occur, we obtain
\begin{equation}
  s_{12}^E \simeq s_{13}^E \simeq \l^2  \ .
\label{s12E}
\end{equation}

 How large is the (1-2) family mixing in the neutrino sector?
In order to answer this question, we  discuss the  texture
 in eq.(\ref{typical}) as studied by Vissani \cite{Vissani}:
%%%%%%%%%%%%%%%%%%%%%%%%%%%%%
\begin{eqnarray}
 M_\n\sim   \left( \matrix{\e_{11}^2 & \e_{12} & \e_{13} \cr
           \e_{12} & 1 & 1 \cr  \e_{13} & 1 & 1 \cr  } \right) \ ,
\label{typicalX}
\end{eqnarray}
%%%%%%%%%%%%%%%%%%%%%%%%%%%%
 \noindent where $\e_{ij}$'s  are the same order except for 
factors of order one. 
  At first, diagonalizing the  (2-3) submatrix of this matrix  by
$s_{23}^\nu=1/\sqrt{2}$,
we get the mass matrix in the new basis:

\begin{eqnarray}
 \bar M_\n\sim   \left( \matrix{\bar\e_{11}^2 & \bar\e_{12} & \bar\e_{13} \cr
     \bar\e_{12} & \delta  & 0 \cr  \bar\e_{13} & 0 & 1 \cr  } \right) \ ,
\label{mass2}
\end{eqnarray} 
\noindent where 
\begin{equation}
      \bar\e_{11}^2=\frac{1}{2}\e_{11}^2 \ ,
 \qquad \bar\e_{12}=\frac{\e_{12}-\e_{13}}{2 \sqrt{2}}\ , \qquad
          \bar\e_{13}=\frac{\e_{12}+\e_{13}}{2 \sqrt{2}} \ , \qquad
          \delta=\frac{m_{\nu_2}}{m_{\nu_3}} \ . 
\label{epsilon}
\end{equation}
\noindent
The mass matrix  in eq.(\ref{mass2}) gives 
\begin{eqnarray}
\sin^2 2\th_{\odot}&=& \frac{4\bar\e_{12}^2}
{(\delta-\bar\e_{11}^2)^2 + 4 \bar \e_{12}^2} \ , \nonumber \\
\nonumber \\
\Delta m_{\odot}^2&=& (\delta + \bar\e_{11}^2)
\sqrt{(\delta - \bar\e_{11}^2)^2 + 4\bar\e_{12}^2} \  \Delta m_{\rm atm}^2 \ ,
\end{eqnarray}
\noindent where $\th_{12}=\th_{\odot}$ is taken.
After eliminating the unknown parameter $\delta$, we obtain

\begin{equation}
 \frac{\Delta m_{\odot}^2}{\Delta m_{\rm atm}^2}
 = 4 \frac{\bar\e_{12}^2}{\sin^2 2\th_{\odot}}(\cos 2\th_{\odot}
\pm \frac{\bar\e_{11}^2}{\bar\e_{12}}\sin 2\th_{\odot} )  \ ,
\label{mastereq}
\end{equation}
\noindent where we take account of  the relative phase $\pm$
 in the coefficients of order one.
Since $\e_{12}$ and $\e_{13}$ are complex in general,
$\bar\e_{12}$ is expected to be comparable to them from 
 eq.(\ref{epsilon}).

%%%%%%%%%%%%%%%%%%%%%%%%%%%%%%%%%%%%%%%%%%%%%%
Let us consider the case of $\e_{12}\simeq \e_{13}\simeq \l^2$,
which corresponds to  the model in eq.(\ref{texture1}),
and so gives  $\bar\e_{11}\simeq \bar\e_{12}\simeq \l^2$.
 We show the allowed region
on the $\sin^2 \th_{\odot}-\Delta m_{\odot}^2$ plane in Fig.1, where
we take  $\bar\e_{11}=\bar\e_{12}=0.05\sim 0.08$ \cite{SY}
and $\Delta m^2_{\rm atm}=(1.5\sim 5)\times 10^{-3} \eV^2$.
In order to show the allowed region in the dark side \cite{dark},
 we use  $\sin^2\th_{\odot}$ as the horizontal axis.
There are two allowed regions corresponding to the sign in the right
hand side of eq.(\ref{mastereq}).
The region between two thick solid curves is allowed in the case of the plus
sign.
The case of the minus sign is shown by the dashed curves.
The tips in the plot  extend down to $\Delta m^2_{\odot}=0$. 
The parameter regions of the LMA-MSW solution (upper) and the LOW solution 
(lower) are shown loosely by the black rectangles at the $95\%$
confidence level in Fig.1.
The region of the SMA-MSW solution  is squeezed in the vertical axis.
The region of the VO solution is omitted.

 As seen in Fig.1, 
our mass matrices are clearly consistent with the LMA-MSW solution.
They may be also consistent with the LOW and VO solutions
because the LOW solution hits the maximal mixing and extends into the 
dark side at the $99\%$ confidence level, and the VO solution is even
more so.

%%%%%%%%%%%%%%%%%%%%%%%%%%%%%%%%%%%%%%%%%%%%%%%%%%%%%%%%
However, if the phase of  $\e_{12}$ is equal to  the one of  $\e_{13}$, 
the parameter $\bar\e_{12}$ could  be supressed in eq.(\ref{epsilon}).  
When we take  $\bar\e_{11}\simeq \l^2$ and $\bar\e_{12}\ll \l^2$, 
 we can get another allowed region. In Fig.1 
 we show the case of  the minus sign in eq.(\ref{mastereq})
 with  typical values $\bar\e_{11}=0.06$ and $\bar\e_{12}=0.003$, and  
 within $\Delta m^2_{\rm atm}=(1.5\sim 5)\times 10^{-3} \eV^2$. 
The region of the case with the plus sign is symmetric respect to
the axis at  $\sin^2\th_{\odot}=0.5$.
  This case  is  consistent with both the SMA-MSW  
 and the LOW solutions
\footnote{In our  calculation, we take account of   $s_{12}^E$
  with  the condition $\a\simeq \pm\pi/2$, which is required to keep
the maximal MNS mixing as seen in eq.(\ref{max}).}.
Thus, if $\bar\e_{12}$ is suppressed by the large cancellation,
the our mass matrices could be  clearly consistent with the SMA-MSW and the
  LOW solutions.

%%%%%%%%%%%%%%%%%%%%%%%%%%%%%%

It is  important  to discuss the magnitude of  $U_{e3}$.
Since  $U_{e3}$ is given approximately as 
$\max \{s_{12}^E, \ s_{13}^E,\ s_{13}^\nu \}$,
\footnote{As far as  $\a\simeq \pm\pi/2$, the cancellation
 never  occur among   $s_{12}^E$, $s_{13}^E$ and $\ s_{13}^\nu$.} 
its expected magnitude  is $\l^2$, which is consistent with
the CHOOZ data \cite{CHOOZ}.

Since the parameter $\delta$ is give as
\begin{equation}
\delta \simeq  2\bar\e_{12}\cot 2\theta_{\odot}+\bar\e_{11}^2 \ ,
\end{equation}
\noindent
the LMA-MSW  solution  
corresponds to  $\delta\simeq \l^2$. 
However,
 there is no  principle to fix  $\delta$
 in advance  within the framework of the $U(1)$ flavor symmetry. 
In other words, the lepton mass matrices in eq.(\ref{texture1}) 
are consitent with the LMA-MSW, SMA-MSW and  LOW   solutions.
They may be consistent even with the VO solution.
Thus, these mass matrices cannot predict the solar neutrino solution
because $\delta$ is the unknown parameter.

%%%%%%%%%%%%%%%%%%%%%%%%%%%%%%%%%%%%%%%%%%%%%%%%%%%%%
%%%%%%%%%%%%%%%%%%%%%%%%%%%%%%%%%%%%%%%%%%%%%%%%%%%%%
Now 
we  go beyond the $U(1)$ flavor symmetry to give the unique solution
of the solar neutrino oscillation.
Let us consider another mass matrix in  $U(1)\times Z_2$ symmetry,
which has already discussed in ref.\cite{Grossman}.
 The effective Yukawa couplings of the lepton sector
 are given by extending eq.(\ref{effectiveL}) with the singlet  Higgs
  $S_1$ and  $S_2$ as follows \cite{Leurer}:

  \begin{equation}         
    L_i \bar \ell_j H_d \e_1^{m_{ij}} \e_2^{m'_{ij}} +
    \frac{1}{M_R} L_i  L_j H_u H_u \e_1^{n_{ij}} \e_2^{n'_{ij}}\ ,
  \end{equation}

  \noindent
 where $\e_1\equiv <S_1>/\Lambda$ and $\e_2\equiv <S_2>/\Lambda$ are
assumed to be  expressed in terms of $\l$.  

%%%%%%%%%%%%%%%%%%%%%%%%%%%%%%%%%%%%%%%%

Giving the $U(1)$ and $Z_2$ charges to  the doublet leptons $L_i$ and 
the singlets  $\bar \ell_j$ as follows:

\begin{equation}      
L_1(1,\ 0), \  L_2(1,\ 0),   L_3(0,\ 1);   \quad\qquad
\ell^c_1(5,\ 0), \  \ell^c_2(2,\ 0), \ \ell^c_3(0,\ 0) ,
\label{U1charge}
\end{equation}

\noindent we obtain

\begin{eqnarray}
 M_E\sim
             \left (\matrix{ e_1\e_1^6 & f_1\e_1^3 & g_1\e_1 \cr
			e_2\e_1^6 & f_2\e_1^3 & g_2\e_1 \    \cr
                e_3\e_1^5 \e_2& f_3\e_1^2\e_2 & g_3 \e_2\    \cr } \right)  ,
  \qquad
 M_\n\sim   \left( \matrix{\e_1^2 & \e_1^2 & \e_1\e_2 \cr
                           \e_1^2 & \e_1^2 & \e_1\e_2 \cr
                           \e_1\e_2 & \e_1\e_2 & 1 \cr  } \right) \ ,
\end{eqnarray}

\noindent
where $e_i$, $f_i$ and $g_i$ are supposed to be real  coefficients of 
order one.
Taking  $\e_1= \e_2=\l$, which is  realized in the supersymmetric
vacuum  \cite{U1X}, we obtain
 $m_{\nu_2}/m_{\nu_3}=\l^2$, which is consistent with the mass scale of
the LMA-MSW solution.  
It is found that the large (2-3) family mixing of the MNS matrix
originates  from $M_E$.  On the other hand,
the large (1-2) family mixing comes from both  $M_E$ and $M_\nu$.
In these lepton mass matrices,
an important problem is the magnitude of  $U_{e3}$.
In order to investigates  $U_{e3}$, we make  $M_E M_E^\dagger$

\begin{eqnarray}
 M_E M_E^\dagger \sim && \l^2\left(\matrix{g_1\cr g_2\cr g_3\cr}\right )
 \left(\matrix{g_1 & g_2  & g_3\cr}\right )
+\l^6  \left(\matrix{f_1\cr f_2\cr f_3\cr}\right )
 \left(\matrix{f_1 & f_2  & f_3\cr}\right ) \nonumber \\
  &&  +  \l^{12} \left(\matrix{e_1\cr e_2\cr e_3\cr}\right )
 \left(\matrix{e_1 & e_2  & e_3\cr}\right )  \ ,
\end{eqnarray}

\noindent
which is the sum of three rank-one matrices.
In general, one cannot expect the small $U_{e3}$, which is given 
approximately as 
\begin{equation}
 |U_{e3}|\simeq\frac{1}{k} \left |\frac{g_1}{g_2} -\frac{f_1}{f_2}\right | \ ,
\end{equation}
with
\begin{equation}
k^2=\frac{1}{f_2^2 g_2^2}\left [
 (f_2 g_3 - f_3 g_2)^2 +(f_1 g_3 - f_3 g_1)^2+(f_2 g_1- f_1 g_2)^2
 \right ] \  ,
 \end{equation}
\noindent 
which is of order one.
Unless  the following relation is kept:
\begin{equation}
\frac{g_1}{g_2}= \frac{f_1}{f_2} \quad {\rm and} \quad
 \frac{g_2}{g_3} \not= \frac{f_2}{f_3} \quad {\rm or} 
\quad \frac{g_1}{g_3} \not= \frac{f_1}{f_3} \ ,
\end{equation}
\noindent
 $U_{e3}$ is not suppressed.
  We have checked numerically that the predicted value of  $U_{e3}$
 is below  the CHOOZ bound  ($U_{e3}<0.16$)
if  these relations are satisfied with the accuracy of  $\sim 10\%$.

Since these coefficients are not constrained by the symmetry,
we cannot expect very  small  $U_{e3}$ in general.
In other words, our mass matrices give the  non-negligible $U_{e3}$.
If  the upper bound of $U_{e3}$ is more constrained by future experiments,
the likelihood of this texture decreases.
Thus,  our  lepton mass matrices, which lead to the LMA-MSW
solution, will be crucially tested by  the measurement of  $U_{e3}$. 

%%%%%%%%%%%%%%%%%%%%%%%%%%%%%%%%%%%%%%%%%%%
%%%%%%%%%%%%%%  Quark Sector %%%%%%%%%%%%%%%
It is helpful to comment on the quark sector.
If the $U(1)\times Z_2$ charges in eq.(\ref{U1charge}) are assigned
in SU(5) as well as in eq.(\ref{SU5}), we get the quark mass matrices:
\begin{eqnarray}
 M_U\sim   \left( \matrix{\l^{10} & \l^7 & \l^5 \cr
                           \l^7 & \l^4 & \l^2 \cr
                           \l^5 & \l^2 & 1 \cr  } \right) \ , \qquad\quad
 M_D\sim
             \left (\matrix{\l^5 & \l^5 & \l^5 \cr
			    \l^2 & \l^2 & \l^2    \cr
                            1 & 1 & 1    \cr } \right) \ . 
\end{eqnarray}
\noindent
These mass matrices are not consistent with the experimental values
of the quark mixing matrix, because these predict
 $V_{us}\simeq \l^3$ and $V_{ub}\simeq \l^5$.  We need some modification
in the first family quark sector.

%%%%%%%%%%%%%%%%%%%%%%%%%%%%%%%%%%%%%%%%%
%%%%%%%%%%%%%%%  Summary %%%%%%%%%%%%%%%%
%%%%%%%%%%%%%%%%%%%%%%%%%%%%%%%%%%%%%%%%%

We have discussed the lepton mass matrices with the $U(1)$ flavor symmetry.
It is found that the solar neutrino solution depends on the parameters of the
next-leading terms such as $\delta$ and $\bar\e_{12}$.
In order to get the unique solution of the LMA-MSW solution,
we have studied the lepton mass matrices  
with the $U(1)\times Z_2$ symmetry.
The coefficients of the matrix elements of the charged leptons
 are  constrained strongly due to  the small $U_{e3}$ in the CHOOZ
experiment.

It should be noted that our results are not spoiled by the
radiative corrections, because three  neutrino masses are  hierarchical,
and moreover,  $\tan\b$ is arbitrary in our models \cite{RGE}.

Since $U_{e3}$ is expected to be seizable in both models, the prediction
will be  tested  in the long baseline experiments in the future.
If  the future experiments  constrain $U_{e3}$ to be  smaller than
$\l^2$, there is no hope to explain the LMA-MSW solution
in the $U(1)$ flavor symmetry.
Thus, the solar neutrino solution makes a big impact on the
lepton mass matrix with the flavor symmetry.
We expect that the LMA-MSW solution  will be tested  in KamLAND
as well as SNO in the near future.

\vskip 0.5  cm
%%%%%%%%%%%%%%%%%%%%%%%%%%%%%%%%%%%%%%%%%%%%%%%%%%%%%
{\large\bf Acknowledgements:}
 We would like to thank  F. Vissani, J. Sato, H. Nakano and
M. Bando  for useful discussions.
 This research is  supported by the Grant-in-Aid for Science Research,
 Ministry of Education, Science and Culture, Japan(No.10640274,
 No.12047220).  

\newpage
%%%%%%%%%%%%%%%%%%%%%%%%%%%%%%%%%%%%%%%%%%%%%%%%%%%%
%%%%%%%%%%%%%%%%%%%%%%%%%%%%%%%%%%%%%%%%%%%%%%%%%%%

\newpage
%%%%%%%%%%%%%%%  Figure 1  %%%%%%%%%%%%%%%%%
\begin{figure}
\epsfxsize=14 cm
\centerline{\epsfbox{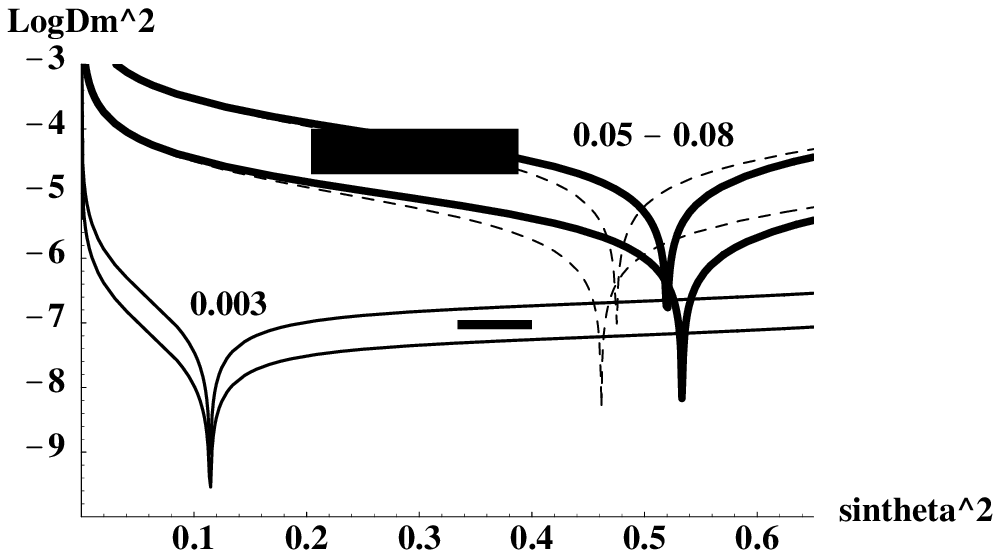}}
\caption{}
\end{figure}
Fig.1: The allowed regions on the
 $\sin^2 \th_{\odot}-\log_{10}(\Delta m_{\odot}^2/\eV^2)$  plane.
 The region between the thick solid curves (the dashed curves)
 corresponds to plus (minus) sign with $\bar\e_{12}=0.05\sim 0.08$.
  The region between thin solid  curves corresponds  to minus sign 
 with $\bar\e_{12}=0.003$.
 Tips in the plot  extend down to $\Delta m^2_{\odot}=0$. 
 The black rectangle regions show the LMA-MSW (upper) and the LOW
(lower) solutions approximately. The region of the SMA-MSW solution
 is squeezed in the  vertical axis.


\begin{thebibliography}{1}
\bibitem{phase}
 For example, if the $M_E(2,3)$ component has the phase $e^{i\a}$,
 the phase matrix appears in eq.(\ref{max}).
 See eqs.(15)-(17) in ref.\cite{HR} for the quark sector.
\end{thebibliography}
\end{document}